\documentclass[final]{ustcstep}

\usepackage[dvips]{graphicx}
\usepackage[square]{natbib}
\usepackage{enumerate}
\def\gsim{\;\lower4pt\hbox{${\buildrel\displaystyle >\over\sim}$}\;}
\def\lsim{\;\lower4pt\hbox{${\buildrel\displaystyle <\over\sim}$}\;}
\def\grls{\;\lower4pt\hbox{${\buildrel\displaystyle >\over <}$}\;}

\newcommand\addr[2]{{\footnotesize \it $^{#1}$#2}\\}
\usepackage{color}           % For color text: \color command
\usepackage{url}             % For breaking URLs easily trough lines
\usepackage{enumerate}
\usepackage[pdfborder={0 0 0},urlcolor=blue,breaklinks]{hyperref}
\ifx \doiurl \undefined \def \doiurl#1{\href{http://dx.doi.org/#1}{\url{#1}}}\fi
\ifx \adsurl \undefined \def \adsurl#1{\href{http://adsabs.harvard.edu/abs/#1}{\url{#1}}}\fi

\newcommand{\adv}{    {\it Adv. Space Res.}} 
 
\newcommand{\aap}{    {\it Astron. Astrophys.}}

\newcommand{\apj}{    {\it Astrophys. J.}}

\newcommand{\apjs}{   {\it Astrophys. J. Suppl.}}

\newcommand{\grl}{    {\it Geophys. Res. Lett.}}

\newcommand{\jgr}{    {\it J. Geophys. Res.}}

\newcommand{\solphys}{{\it Solar Phys.}}
 
\newcommand{\ssr}{    {\it Space Sci. Rev.}}

\begin{document}

\title{Source Region of the Decameter--Hectometric Type II Radio Burst: Shock--Streamer Interaction Region}
\author{Chenglong Shen, Chijian Liao, Yuming  Wang, Pinzhong Ye, Shui Wang \\ 
        \addr{}{CAS Key Laboratory of Geospace Environment, 
	Department of Geophysics and Planetary Sciences,} 
	\addr{}{University of Science and Technology of China, Hefei, Anhui 230026, China}
    \addr{}{\href{mailto:clshen@ustc.edu.cn}{clshen@ustc.edu.cn}}}
\maketitle
\tableofcontents

\begin{abstract}
D--H type II radio bursts are widely thought to be caused by the coronal mass ejections (CMEs).
However, it is still unclear where the exact source of the type IIs on the shock surface is. 
We identify the source regions of the decameter--hectometric (D--H) type IIs based on 
imaging observations from SOHO/LASCO and the radio dynamic spectrum from \textit{Wind/Waves}. 
The analysis of two well--observed events suggests that the sources of these two events are 
located in the interaction regions between shocks and streamers, and that
the shocks are enhanced significantly in these regions.
\end{abstract}
%-------------------------------------------------

\section{Introduction}
\label{Introduction} 

Type II radio bursts, especially in the decameter--hectometric (D--H) and kilometer (km) wavelength range, 
are thought to be caused by the electron beam accelerated by CME--driven shocks
\citep[e.g.][]{SheeleyJr:1985taa,1998GeoRL..25.2493R,1999GeoRL..26.1573B}.
Assuming that the type II radio burst was excited at the shock front,
the speed of the shock could be obtained from the observed frequency drift rate of the type II radio burst
based on a coronal--density model \citep[e.g.][]{2001A&A...377..321V,Gopalswamy:2002jq,2002A&A...396..673V,2003AIPC..679..152R,2004A&A...413..753V}. 
This method is widely used to study and forecast the propagation of shocks  
\citep[e.g.][]{Dryer:1984ua,1990SoPh..129..387S,Fry:2003vh,2007ApJ...663.1369R}.
\citet{Shen:2007ww} established a method to derive shock strength from the observations 
based on the assumption that type II radio bursts were generated from the nose of shocks. 

However, whether or not type II radio bursts originate from the nose of shocks is still an open question.
As a high--density region, the streamer was thought to be a place where strong shocks easily form \citep{2008ApJ...687.1355E}. 
Thus, the shock--streamer interaction region was discussed as a possible source region of type II radio bursts
\citep[e.g.][]{Cho:2005jv,Cho:2007da,Cho:2008cf,Cho:2011be}. 
Using the coronal--density distribution obtained from the Mauna Loa Solar Observatory ( MLSO )
MK4 polarization map, \citet{Cho:2007da} studied the relationship between a metric type II burst and a CME. 
They found that the metric type II burst was generated at the interface of the CME flank and the streamer. 
Further, \citet{Cho:2008cf} checked the source regions of 19 metric type II radio 
bursts and found that both of the front and the CME--streamer interaction regions are the possible source regions for the metric type II radio bursts.
In addition, the shock and streamer interaction could also affect the spectrum of the type II radio burst.
Recently, \citet{2012ApJ...750..158K} and \citet{2012ApJ...753...21F} reported two clear cases 
where the spectrum of type II radio bursts varied during the interaction between the shocks and the streamers. 

%Previous results show that the interaction region between the shock and streamer was the possible source region of metric type II radio burst. 
However, the discussion of the source regions of the type II radio bursts 
in the decameter--hectometric (D--H) wavelength range is still lacking. 
The spectrum of RAD2 on \textit{Wind/Waves} \citep{Bougeret:1995ce} is from 1 to 14 MHz 
corresponding to the corona region from $\approx$2 $R_\odot$ to $\approx$9 $R_\odot$.
The C2 camera of the \textit{Large Angle and Spectrometric COronagraph} (LASCO; \citet{Brueckner:1995cb})
onboard the \textit{SOlar and Heliospheric Observatory} (SOHO; \citet{Domingo:1995}) 
provides the imaging observations from 1.5 $R_\odot$ to 6 $R_\odot$.
Thus, the combination of \textit{Wind/Waves} and SOHO/LASCO--C2 observations could be used 
to study the source regions and the variation of type II radio bursts at the D--H wavelength range. 
In this article, we check source regions of two well--observed D--H type II radio bursts.
The method applied in this study is introduced in Section 2.
In Sections 3 and 4, two typical D--H type II events and their source regions are studied. 
Conclusions and discussions are then given in the last section.

\section{Method}\label{method}

In this work, the source region of the D--H type II radio burst is obtained 
from the combined analysis of the \textit{Wind/Waves} and SOHO/LASCO observations. 
The detailed method is described as follows: 

\begin{enumerate}[i)]

\item Previous results suggest that type II radio bursts especially 
in the D--H and longer wavelength range are caused by the CME--driven shocks \citep{1998GeoRL..25.2493R,1999GeoRL..26.1573B}. 
Recently, \citet{Vourlidas:2003kn} and \citet{2009ApJ...693..267O} found 
that shocks could be directly observed in coronagraph images.
We use SOHO/LASCO--C2 observations to identify the position of the shock front, 
called S$_{\mbox{shock}}$ hereafter. It is thought to be the possible source region of type II bursts.

\item The fundamental frequency of a type II burst is related to 
the background electron density by \citep{1982GAM....21.....P}:
\begin{eqnarray}
N_e=(\frac{f_{pe}(Hz)}{8.98\times10^3})^2\mbox{cm}^{-3}\label{eq1}.
\end{eqnarray} 
Therefore, the electron density of the source regions of the type II could be obtained from the radio burst dynamic spectrum.
Using the pb\_inverter.pro procedure in Solar Software (SSW\footnote{\href{http://www.lmsal.com/solarsoft/}{http://www.lmsal.com/solarsoft/}}), 
the polarized--brightness observations from SOHO/LASCO could be used to get the background electron density distribution. 
The pb\_inverter.pro procedure uses the pB inversion derivation obtained by \citet{1950BAN....11..135V}.
A polynomial fit of the form r$^{-n}$ is applied to the pB image for a single position angle  
to get the electron density distribution (see the introduction of the `pb\_inverter.pro' in the SSW).
%\inlinecite{Hayes:2001ii} discussed this method and its application on total brightness observation detailed.}
Thus, the possible regions, called as S$_\rho$, which can generate the type II radio bursts
at the observed frequency range at the time of shock observed,
can be determined. 
In addition, considering a 2\% uncertainty in the brightness observations (Vourlidas, 2012, private communicate), 
a 2\% uncertainty in the obtained electron density was applied to find the S$_\rho$. 

\item The overlap region of the shock front (S$_{\mbox{shock}}$)
and the derived density region (S$_\rho$) is defined as the source region of the type II radio burst.
\end{enumerate}

Based on the method described above, to determine the source region of a D--H type II radio burst, 
we need the polarized--brightness image and the direct imaging observations of the shock from SOHO/LASCO 
and the D--H type II radio--burst observation from \textit{Wind/Waves}. 
Thus, we select events based on the following criteria:

\begin{enumerate}[i)]
\item A clear type II radio burst was recorded by \textit{Wind/Waves}. 

\item The burst was caused by a limb CME with clear shock signatures in the LASCO--C2 field of view.  
We require limb CMEs because the polarized--brightness image represents the background 
coronal--density distribution near the plane of the sky, and the shock driven by a limb CME should have 
the most clear signatures in coronagraph.
\end{enumerate}

Conforming to these two criteria, two well--observed  events were found for study in this article.

\section{7 March 2011 Event}

\begin{figure}
\center
 \noindent\includegraphics[width=\hsize]{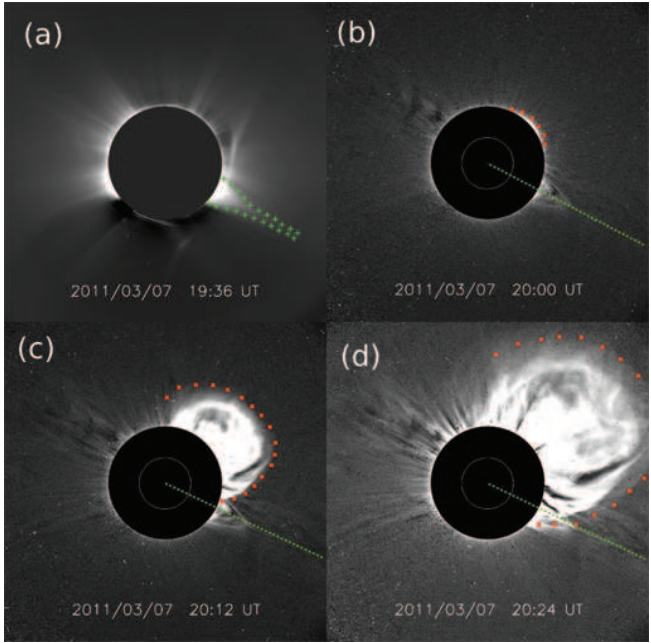}
 \caption{The SOHO/LASCO--C2 observations for the 7 March 2011 event. Panel (a) shows the polarized--brightness image. 
Panels (b) -- (d) show the CME at different times.} 
\label{mar7_lasco}
 \end{figure}

Figure \ref{mar7_lasco} shows the SOHO/LASCO observations before and after the onset of this CME.
On 7 March 2011, a limb CME originating from N24W59 was first observed by SOHO/LASCO--C2 at 20:00 UT. 
The orange $\ast$ symbols in panel (b) of Figure \ref{mar7_lasco} show the possible front positions of this CME at 20:00 UT.
Using the GCS model \citep{Thernisien:2006ke,2009SoPh..256..111T,Thernisien:2011jy}, 
\citet{chenglong:2012wa} obtained the speed of this CME as 2115 $\pm$ 136 km\ s$^{-1}$ in the three--dimensional space.
This is a very fast CME with a speed much faster than the local Alfv\'en speed, and therefore LASCO--C2 only captured three images of the CME.
We can expect that this CME drove a shock when it propagated in the corona.

Seen from panel (c) and (d) in Figure \ref{mar7_lasco}, obvious shock signatures ahead of the main body of the CME
could be identified.
The orange $\ast$ symbols in panel (c) and (d) of Figure \ref{mar7_lasco} 
mark the shock front at two instants of time.
As we described in Section 2, the shock front, S$_{\mbox{shock}}$, is thought to be 
the possible source region of the associated type II radio burst. 
 
\begin{figure}
\center
 \noindent\includegraphics[width=\hsize]{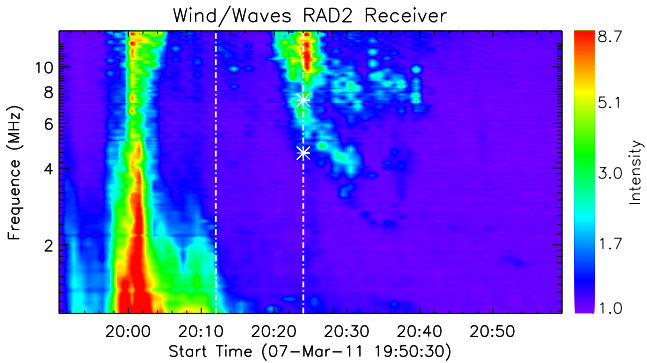}
 \caption{The \textit{Wind/Waves} observations of 7 March 2011 19:50 UT to 21:00 UT.} 
%The vertical dashed lines denote the time that shocks were observed by SOHO/LASCO C2. 
%The white `$\ast$' symbols show the minimum and maximum value of the type II radio burst' fundamental frequency. }
\label{mar7_wave}
 \end{figure}

Figure \ref{mar7_wave} shows the \textit{Wind/Waves} observations from 7 March 2011 19:50 UT to 21:00 UT. 
From Figure \ref{mar7_wave}, an obvious type II radio burst could be identified. 
The vertical dotted--dashed white lines in Figure \ref{mar7_wave} indicate the times of the shock recorded by SOHO/LASCO--C2.
Seen from this figure, the signature of the type II radio burst at 20:12 UT is very weak. 
Near 20:22 UT, this type II radio burst became stronger, and lasted for about ten minutes.  
The white asterisks show the minimum and maximum fundamental frequency of this D--H type II radio burst at the time of 20:24 UT, 
which are 4.6 and 7.4 MHz corresponding to the electron density of $2.6\times10^5$ cm$^{-3}$ and $6.8\times10^5$ cm$^{-3}$, 
respectively, based on Equation \ref{eq1}. 

\begin{figure}
\center
 \noindent\includegraphics[width=\hsize]{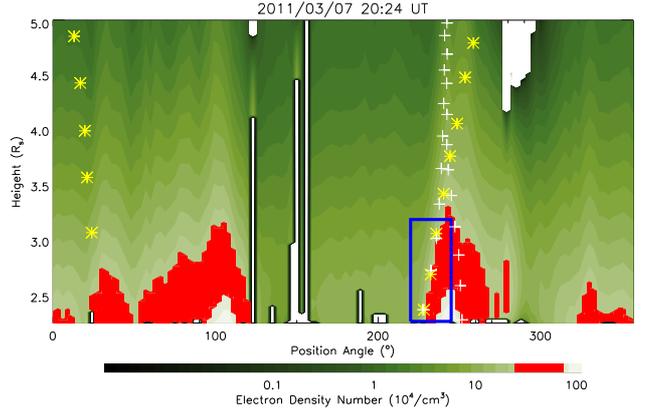}
 \caption{The electron--density distribution obtained from the polarized--brightness image from SOHO/LASCO for the 7 March 2011 event. 
}\label{mar7_rho}
 \end{figure}

Figure \ref{mar7_rho} shows the background electron--density distribution obtained from the polarized image 
at 08:58 UT. 
The white regions in Figure \ref{mar7_rho} are caused by the unsuccessful determination of the density.
It is clearly seen in the figure that the electron density varies significantly with position angle.
This suggests that a simple one--dimensional density model may not reflect the real condition. 
The  regions, i.e. S$_\rho$, in which the electron density falls in the range from $2.6\times10^5$ to $6.8\times 10^5$ cm$^{-3}$, 
are indicated in red. The type II radio burst near 20:24 UT were probably generated from these region.

The position of S$_{\mbox{shock}}$ based on coronagraph observations is overplotted with yellow $\ast$ on Figure 3.   
As we discussed in Section \ref{method}, the source region of a 
type II radio burst is the overlap region between the S$_{\mbox{shock}}$ and S$_\rho$.
For this event at 7 March 2011 20:24 UT, 
the source region is located in the region indicated by the blue rectangle.

The white + symbols in Figure \ref{mar7_rho} show the boundary of the streamer,
which is determined from the SOHO/LASCO image before the onset of the CME as indicated by the green + in Figure \ref{mar7_lasco}(a). 
It is found that the source of this type II radio burst is located at the shock--streamer interaction region.
This result suggests that the type II radio burst at the D--H frequency range might also originate from the 
shock--streamer interaction region, similar to the metric type II radio bursts \citep{Cho:2007da,Cho:2008cf}.

In addition, at 20:12 UT, the shock was also very clear in the SOHO/LASCO--C2 image (Figure 1c),
but, the type II signature was much weaker than that near 20:24 UT.
We suggest that such a difference is probably attributed to the degree of interaction between the shock and the streamer. 
From Figure \ref{mar7_lasco}(c),
it seems that the shock did not fully interact with the streamer at 20:12 UT. 
But, at 20:24 UT as shown in Figure \ref{mar7_lasco}(d), 
a part of the CME--driven shock was propagating in  the streamer obviously. 
The interaction between the streamer and the shock enhanced the shock and 
then the enhanced shock increased the intensity of the type II radio burst.

\section{9 May 2011 event}
\begin{figure}
\center
 \noindent\includegraphics[width=\hsize]{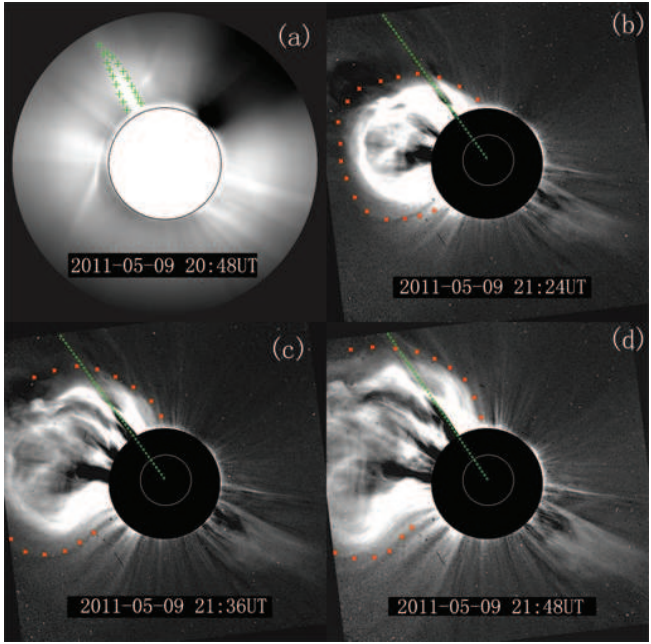}
 \caption{The SOHO/LASCO--C2 observations for the 9 May 2011 event as for Figure \ref{mar7_lasco}. 
}\label{may9_lasco}
 \end{figure}

This CME burst from the east limb of the solar disk. 
SOHO/LASCO--C2 observed it since 21:24 UT. 
It was a fast limb CME with a projected speed of 1318 km\ s$^{-1}$. 
Figure \ref{may9_lasco} shows the SOHO/LASCO observations before and after the onset of this CME.
It is found that the shock structure ahead of the CME could be well observed and identified 
at 21:24 UT, 21:36 UT and 21:48 UT based on SOHO/LASCO--C2 observations.
The orange $\ast$ symbols in panels (b) -- (d) show the shock front at three instants of time, which are defined as S$_{\mbox{shock}}$.

\begin{figure}
\center
 \noindent\includegraphics[width=\hsize]{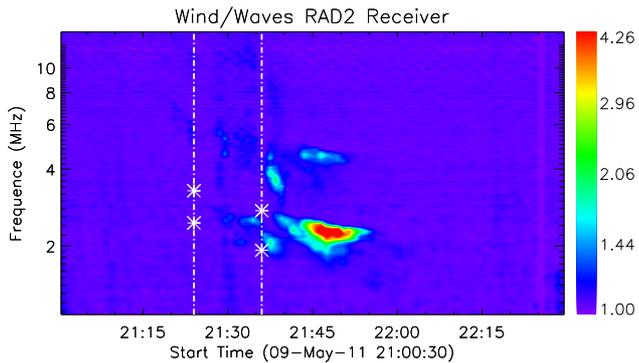}
 \caption{The \textit{Wind/Waves} observations from 9 May 2011 21:00 UT to 22:30 UT, similar to Figure \ref{mar7_wave}.}\label{may9_wave}
 \end{figure}

A D--H type II radio burst associated with this CME is shown in Figure \ref{may9_wave}.
It started at $\approx$21:15 UT.
At 21:24 UT, the type II radio burst was very weak and is difficult to identify.
The half of the frequency of its harmonic component, which varied from 2.5 MHz to 3.3 MHz, is used as the fundamental frequency. 
After 21:24 UT, the strength of this D--H type II increased. 
This radio burst reached its strongest phase near 21:48 UT.
At 21:36 UT, the fundamental frequency varied from 1.9 MHz to 2.7 MHz.

\begin{figure}
\center
 \noindent\includegraphics[width=\hsize]{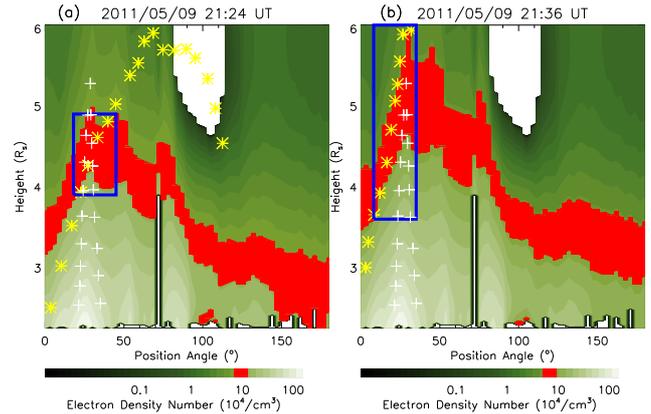}
 \caption{ The electron--density distribution obtained from the polarized--brightness image from SOHO/LASCO for the 9 May 2011 event. 
}\label{may9_rho}
 \end{figure}

Figure \ref{may9_rho} shows the electron--density distribution, which is derived 
from the polarized--brightness image recorded at 14:58 UT. 
The regions of S$_\rho$ at  21:24 UT and 21:36 UT are indicated by the red color in panels (a) and (b),  respectively.
Similar to Figure \ref{mar7_rho}, the yellow $\ast$ symbols
indicate S$_{\mbox{shock}}$. 
The source regions of this type II event at the two instants of time were located in the regions 
enclosed by the blue rectangles in Figure \ref{may9_rho}.

The white + symbols in Figure \ref{may9_rho} show the boundary of the streamer as same as the green + in Figure \ref{may9_lasco}(a).
It is found that the source regions of this event at different time also located in the shock--streamer interaction regions. 
It confirms the conclusion that the shock--streamer interaction region might be the source region of a D--H type II radio burst.
Seen from the Figure \ref{may9_lasco} (b) and (c), it is found that the interaction 
between the shock and the streamer may start near 21:24 UT. 
After that, the shock further interacted with the streamer.
During this phase, the observed D--H type II radio burst enhanced continuously as shown in Figure \ref{may9_wave}. 
Thus, the increase of the intensity of this radio burst was probably caused by the enhancement of the shock during its interaction with the streamer. 


\section{Conclusion}
In this work, the source regions of two well--observed D--H type II radio bursts 
are checked based on SOHO/LASCO--C2 and \textit{Wind/Waves} observations. 
It is found that the source regions of these two D--H type II radio bursts probably located 
in the shock--streamer interaction regions, which is the same as the source regions of metric type II radio bursts \citep{Cho:2007da,Cho:2008cf,Cho:2011be}. 
In addition, by analyzing the intensity variation of these two D--H type II radio bursts, 
we suggest that the shocks were enhanced during their interaction with the streamer.
Such enhancement of shocks would increase the intensity of the radio bursts.

These results indicate that the shock--streamer interaction region 
could also be one of the main source region of the D--H type II radio burst.
It should be noted that the background density in a streamer (or the flank of a shock) 
is quite different from that near the nose of a shock. 
Thus, to calculate the shock speed based on the frequency drift rate of type II radio burst, 
a detailed analysis of where a radio burst is generated from should be done first.  

As we described in Section \ref{method}, the background density obtained from the SOHO/LASCO 
polarized--brightness image is an important factor in our method. 
Unfortunately, there is only one polarized image taken each day in each telescope 
for most period of the SOHO mission\footnote{\href{http://lasco-www.nrl.navy.mil/index.php?p=content/retrieve/products}{http://lasco-www.nrl.navy.mil/index.php?p=content/retrieve/products}}.
Solar eruptions, especially the large CMEs, would significantly influence the background density.
Thus, we choose only the events in which no large CME events occurred between the time of the polarized--brightness 
image recorded and the type II radio burst.
In addition, clear type II radio--burst observations and clear shock signatures in SOHO/LASCO observations are needed in this method.
Combined with these selection criteria, the number of events could be studied is limited. 
In a future work,  the method developed by \citet{Hayes:2001ii} could be used to obtain the background electron density based on the 
total--brightness images, and more events could be studied.

It should be noted that only projection observations were used in this work. 
Recently, some methods were developed to obtain the CME's parameters \citep[e.g.][]{Thernisien:2006ke,
2009SoPh..256..111T; Thernisien:2011jy; 2012arXiv1203.3261F}, 
background electron density in the corona \citep[e.g.][]{Frazin:2010fc} 
and the streamer structure \citep[e.g.][]{2010ApJ...710....1M} 
in three--dimensional space. 
Thus, the three--dimensional source region of the radio burst could be further checked by applying various three--dimensional models. 

We acknowledge the use the SOHO/LASCO and \textit{Wind/Waves} observations.
The SOHO/LASCO data used here are produced by a consortium of the Naval Research Laboratory (USA), Max-Planck-Institut f\"{u}r Aeronomie (Germany),
Laboratoire d'Astrophysique de Marseille (France), and the University of Birmingham (UK).
SOHO is a mission of international cooperation between ESA and NASA. 
This work is supported by the CAS Key Research Program (KZZD-EW-01),
grants from the 973 key project 2011CB811403, NSFC 41131065, 40904046, 41274173, 40874075, and 41121003,
CAS the 100-talent program, KZCX2-YW-QN511 and startup fund, and MOEC 20113402110001 and the
fundamental research funds for the central universities (WK2080000031 and WK2080000007).

\end{document}